\pdfoutput=1

\documentclass[sigconf]{acmart}

\RequirePackage[l2tabu, orthodox]{nag}

\usepackage{dirtytalk}
\usepackage{siunitx}
\usepackage{listings}
\usepackage[inline]{enumitem}
\usepackage{cleveref}

\usepackage[T1]{fontenc}
\usepackage[latin9]{inputenc}

\newcommand{\co}{compiler}
\newcommand{\cf}{\co\ flag}
\newcommand{\clang}{{C}lang}
\newcommand{\gcc}{\textsc{GCC}}
\newcommand{\msvc}{\textsc{MSVC}}

\acmConference[ICSE 2024]{46th International Conference on Software Engineering}{April 2024}{Lisbon, Portugal}

\title{The Devil Is in the Command Line: Associating the Compiler Flags With the Binary and Build Metadata}

\begin{abstract}
Engineers build large software systems for multiple architectures, operating systems, and configurations.
A set of inconsistent or missing compiler flags generates code that catastrophically impacts the system's behavior.
In the authors' industry experience, defects caused by an undesired combination of compiler flags are common in nontrivial software projects.
We are unaware of any build and \textsc{CI/CD} systems that track how the \co\ produces a specific binary in a structured manner.
We postulate that a queryable database of how the \co\ compiled and linked the software system will help to detect defects earlier and reduce the debugging time.
\end{abstract}

\begin{document}

\author[1]{Gunnar Kudrjavets}
\orcid{0000-0003-3730-4692}
\affiliation[obeypunctuation=true]{
   \institution{University of Groningen\\}
   \city{Groningen, }
   \postcode{9712 CP}
   \country{Netherlands}}
\email{g.kudrjavets@rug.nl}

\author[2]{Aditya Kumar}
\orcid{0000-0001-6312-2898}
\affiliation{
    \institution{Google}
    \streetaddress{1600 Amphitheatre Parkway}
    \city{Mountain View}
    \state{CA}
    \country{USA}
    \postcode{94043}}
\email{appujee@google.com}

\author[3]{Jeff Thomas}
\orcid{0000-0002-8026-9637}
\affiliation{
    \institution{Meta Platforms, Inc.}
    \streetaddress{1 Hacker Way}
    \city{Menlo Park}
    \state{CA}
    \country{USA}
    \postcode{94025}}
\email{jeffdthomas@meta.com}

\author[4]{Ayushi Rastogi}
\orcid{0000-0002-0939-6887}
\affiliation[obeypunctuation=true]{
   \institution{University of Groningen\\}
   \city{Groningen, }
   \postcode{9712 CP}
   \country{Netherlands}}
\email{a.rastogi@rug.nl}



\maketitle

\section{Introduction and background}

Compilers are software systems that translate programs \say{into a form in which it can be executed by a computer}~\cite{aho_compilers_2007}.
A {C} or {C++} \co\ such as \clang, \gcc, or \msvc\ supports hundreds of command-line arguments (flags, options, switches).
The \emph{\cf s} instruct \co\ on different aspects of code generation, types of error detection, compliance to a specific version of the programming language standard, or target platform-specific nuances.
\emph{An incorrect combination of \cf s can have disastrous consequences for the resulting software system}.
For example, accidentally turning off the \cf\ to enable checks for buffer security to catch stack overflows (e.g., omitting the \texttt{/GS} option in \msvc) can expose a zero-day vulnerability~\cite{howard_2002,howard_sdl_2006}.

Engineers can compile the same version of a software system to target different platforms and intents, such as debugging, profiling, or an official release.
Typical implicit variables that influence the final set of \cf s are the host operating system where the \co\ executes, the target operating system where the code will run, the \co\ version, dependencies available on the host system, and the desired build type~\cite{build_master_2005,smith_2011}.

Both commercial and open-source software utilizes a variety of \emph{build systems}.
The build system determines the conditions under which a \co\ runs and what combination of \cf s it passes to the \co.
Some of the most popular build systems are: Ant~\cite{ant}, Bazel~\cite{bazel}, Buck2~\cite{buck}, CMake~\cite{cmake}, \textsc{GNU} Make~\cite{gmake}, Ninja~\cite{ninja}, and \textsc{NMAKE}~\cite{nmake}.
Each build system has different means of specifying the dependency graph, defining the rules for build actions, and how it initializes the default set of \cf s.

\Cref{code:redis-makefile} displays a simple conditional statement that modifies the set of dependent libraries based on the host operating system.\footnote{\url{https://github.com/redis/redis/blob/unstable/src/Makefile}}

\lstset{language=make,breaklines=true,frame=single,basicstyle=\ttfamily\footnotesize,caption={Excerpt from the Redis Makefile.},label=code:redis-makefile}
\begin{lstlisting}
# Linux ARM32 needs -latomic at linking time
ifneq (,$(findstring armv,$(uname_M)))
        FINAL_LIBS+=-latomic
endif
\end{lstlisting}

In~\Cref{code:rocksdb-makefile}, we see a more complex conditional logic.\footnote{\url{https://github.com/facebook/rocksdb/blob/main/Makefile}}
The build system disables the usage of a critical dependency (the jemalloc memory allocator) based on what the host and target platforms are.

\lstset{language=make,breaklines=true,frame=single,basicstyle=\ttfamily\footnotesize,caption={Excerpt from the RocksDB Makefile.},label=code:rocksdb-makefile}
\begin{lstlisting}
ifeq ($(PLATFORM), OS_MACOSX)
ifeq ($(ARCHFLAG), -arch arm64)
ifneq ($(MACHINE), arm64)
  ...
  DISABLE_JEMALLOC=1
  PLATFORM_CCFLAGS := $(filter-out -march=native, ...)
  PLATFORM_CXXFLAGS := $(filter-out -march=native, ...)
endif
endif
endif
\end{lstlisting}

By design, a modern build system lets users define build instructions at a higher abstraction level than the formal \co\ command lines.
In \Cref{code:buck_cpp_binary}, we can see how to define a {C++} executable \texttt{foo} in Buck2~\cite{buck}.
The executable \texttt{foo} contains one source code file and one header file.
It also depends on another {C} or {C++} library called \texttt{bar}.

\lstset{language=make,breaklines=true,frame=single,basicstyle=\ttfamily\footnotesize,caption={Sample Buck2 build definition for a {C++} binary.},label=code:buck_cpp_binary}
\begin{lstlisting}
cxx_binary(
  name = 'foo',
  srcs = [ 'foo.cxx', ],
  headers = [ 'foo.hxx', ],
  deps = [ ':bar', ],
)
\end{lstlisting}

For an engineer to understand the details about how \emph{exactly} the \co\ generates code, they need to either intercept the \co\ execution or inspect the resulting log files with the final command line that the \co\ interprets.

\section{Industry challenges}
\label{sec:ind_challenges}

Our primary motivation for this paper comes from observing, debugging, and fixing the repeating patterns of defects.
The root cause for these defects are the incorrect assumptions about how a \co\ generated binaries for a particular software system.
A possible negative interaction between \cf s is a known problem~\cite{pinkers_2004}.
In our industry experience, \emph{the defects caused by incorrect (extraneous, missing, unsuitable) \cf s have high consequences, are hard to detect and stealthy, and are time-consuming to investigate and replicate}.

The primary categories of problems that we have  encountered during the last two decades in the industry are as follows:

\begin{enumerate}
    \item
    \textbf{The differences between engineers' development environment and the official build servers}.
    The official build servers can run on a different operating system, use a different \co\ version, have different environment variables set, or use a different set of preprocessor directives.

    \item
    \textbf{A lack of formal alerting mechanism when the \cf s change}.
    Any seemingly unrelated commit can influence how the \co\ generates the code or what are the application's dependencies.
    The resulting \cf s can change because of confounding variables, such as  environment settings, \co\ configuration files, or corporate device management policies.

    \item
    \textbf{A lack of tools to detect anomalies in the resulting build}.
    A \co\ can generate a subset of a software system differently than the rest.
    Sample issues include
    \begin{enumerate*}[label=(\alph*),before=\unskip{ }, itemjoin={{, }}, itemjoin*={{, and }}]
        \item
        a release build that targets the production environment includes a component with debug tracing enabled
        \item
        global optimization options do not propagate correctly to all the dependencies
        \item
        a consumer that uses different versions of the same dependency in parallel (some things are worse than the infamous \say{\textsc{DLL} hell}~\cite{dll_hell}).
    \end{enumerate*}

    \item
    \textbf{Inability to easily detect syntactic mistakes}.
    An engineer may mistype a \texttt{CXFLAGS} macro in the makefile instead of \texttt{CXXFLAGS} (extra flags passed to a {C++} \co).
    Most build systems lack the means to detect and notify engineers of these mistakes.

    \item
    \textbf{Third-party software components rarely provide the build configuration used to generate the binaries}.
    For example, a dependency can turn off the support for exception handling (e.g., specifying the \texttt{-fno-exceptions} flag in \gcc).
    If the consumer assumes that it can catch exceptions, it invalidates the application's ability to handle errors.
\end{enumerate}

\subsection{Non-deterministic builds}

Popular {C++} compilers can generate code in a non-deterministic manner.
Recompiling a translation unit using the same build configuration can result in a different binary~\cite{grang_fighting_2018,grang_llvm_2017}.
Similarly, modern build systems process the build graph efficiently by enforcing directed acyclic graph-like build dependencies~\cite{bazel,buck}.
As a result, the order of object files listed during the linking stage will depend on which translation unit the \co\ built first.
That, in turn, can cause non-deterministic behavior~\cite{ld_man}.

Tracking the usage of \cf s will decrease a subset of problems related to the reproducibility of the build environment~\cite{lamb_2022}.
It will \emph{help with debugging and early detection of complex defects that influence the behavior of an entire software system}.

\section{Industry needs}
\label{sec:industry_needs}

While the \say{[b]uild systems are awesome, terrifying---and unloved,} they are something that each engineer uses daily~\cite{mokhov_2018}.
Modern build tools such as Bazel~\cite{bazel} and Buck2~\cite{buck}, have made significant progress towards the build hermeticity~\cite{bazel_hermeticity} and making builds reproducible.
However, developing a fully self-contained and deterministic build system has been a complex problem, even for a company like Microsoft~\cite{build_master_2005,smith_2011,schulte_2016}.
Popular utilities such as {GNU} {Autotools} are not hermetic by design and rely on the dependencies from the current execution environment~\cite{calcote_autotools_2020}.

\emph{Engineers need to have the means to understand the evolution of the final set of \cf s for each binary throughout the project's history for each configuration}.
Currently, the \say{state-of-the-art solution} involves engineers inspecting the build logs as a text file and using tools such as \texttt{diff} to compare the results between two builds.
If the build logs are not systematically archived, engineers must rebuild the entire system to understand the final set of \cf s.
Building multiple product versions to isolate a problem can take days or longer for complex software systems, such as an operating system.

\section{Potential research directions}

An obvious solution is to parse the build logs, extract the necessary information, store it in the metadata associated with each resulting build, and provide a query interface for engineers to solve problems similar to the ones we enumerate in~\Cref{sec:ind_challenges}.
Most industrial {CI/CD} systems enable associating an individual build with the test case results, the state of the source code repository when the \co\ generated the binaries, and other related metadata.
Adding the information about \cf s is just another dimension of the metadata.

Another option is storing the \cf s the build system uses inside each binary.
For example, an \textsc{ELF} file format supports the \texttt{.comment} and \texttt{.note} sections in the final binary~\cite{elf}.
However, that approach will require making changes to each used \co.

\balance

\bibliographystyle{ACM-Reference-Format}
\bibliography{compiler-flags}

\end{document}